\documentclass{aa}  
\usepackage[english]{babel}
\usepackage{amsmath}
\usepackage{graphicx}
\usepackage{natbib}
\usepackage[colorlinks=true, allcolors=blue]{hyperref}
\usepackage{txfonts}
\begin{document}

 \title{Consistency between X-ray and UV-Optical reverberation measurements in NGC~5548
 }
 \titlerunning{Characteristics of X-ray, UV, and optical reverberation signatures in NGC~5548}

   %\subtitle{I. Overviewing the $\kappa$-mechanism}

   \author{Amit Kumar Mandal \inst{1,\thanks{amitastro.am@gmail.com}}
    \and
    Bo\.zena Czerny \inst{1}
    \and
    Michal Dov\v{c}iak \inst{2}
    \and
    Elias Kammoun
    \inst{3}
    \and
    Iossif Papadakis \inst{4,5}
          }

   \institute{Center for Theoretical Physics, Polish Academy of Sciences, Al. Lotnik\'ow 32/46, 02-668 Warsaw, Poland
\and
Astronomical Institute of the Czech Academy of Sciences, Boční-II 1401, Praha 4, Prague, 141 00, Czech Republic
\and
Cahill Center for Astronomy \& Astrophysics, California Institute of Technology, 1216 East California Boulevard, Pasadena, CA 91125, USA
\and
Department of Physics and Institute of Theoretical and Computational Physics, University of Crete, 71003 Heraklion, Crete, Greece
\and
 Institute of Astrophysics, FORTH, GR-71110 Heraklion, Crete, Greece
}

   \date{}
\abstract
   {The hard X-ray–emitting hot corona is a key component of active galaxies. Constraints on the hot corona height can be derived from reverberation studies in both the X-ray and optical bands.}
   {X-ray reverberation (X-ray$-$RM) studies often imply a very low corona height, whereas UV/optical reverberation mapping (photometrcic continuum$-$RM) typically points to a much larger one. To reconcile this discrepancy, we examine the constraints provided by both methods for the same source.}
   {We adopt a uniform methodology using the {\tt KYNSED} and {\tt KYNXiltr} codes within a consistent modeling framework for reverberation mapping, applicable across both the X-ray and UV–optical spectral and time domains. We select the source NGC~5548, for which the necessary observational data are available in the literature.}
   {We carry out our analysis for NGC~5548, a source with extensive reverberation mapping data obtained independently in the X-ray and UV–optical bands across different epochs. Our results  hint for
   % reveal 
   a substantial discrepancy between the global parameters required to reproduce the X-ray and those needed to fit the UV–optical reverberation signals. In particular, the mismatch in the inferred black hole mass and accretion rate presents a significant challenge for interpreting the observed time delays within a unified reflection-based framework.}
   {Our unified reflection-based modeling sheds light on X-ray and UV-optical variability of NGC~5548, but discrepancies in black hole mass, accretion rate, and corona properties 
    might imply
   %reveal 
   fundamental challenges to a self-consistent model.  However, future analyses leveraging extended X-ray dataset with improved treatment of absorption and variability coherence are crucial to obtaining more robust constraints.}

   \keywords{Accretion, accretion disks, Galaxies: active, quasars, reverberation mapping
               }
\maketitle

\section{Introduction}

The fundamental components and geometry of active galactic nuclei (AGNs) are now well constrained \citep[see, e.g.,][for general reviews]{1995PASP..107..803U, krolik_book1999,padovani_rev_2017}. In the standard framework, an AGN consists of a central supermassive black hole (SMBH) surrounded by an accretion disk with a compact hot corona above it, a broad emission line region (BLR), and a dusty or molecular torus. Together, these components produce the characteristic broadband AGN spectrum, in which hard X-ray emission arises from the corona, UV/optical radiation is generated predominantly in the accretion disk, and near-infrared (NIR) originates in the dusty torus. In Type 1 AGNs, all of these elements are directly visible, whereas in Type 2 sources the torus partially obscures the nuclear regions due to the high inclination angle ($i$). In addition, a subset of AGNs exhibits powerful relativistic radio jets that further contribute to the observed emission.

Over the past decades, light-echo studies, collectively known as reverberation mapping  \citep[RM;][]{1982ApJ...255..419B, 1993PASP..105..247P} have not only confirmed this structural picture but have also provided a powerful means of probing the innermost regions of AGNs in detail  \citep[see][for a recent review]{Cacket_Sci2021}. By exploiting time variability, these techniques effectively use temporal resolution for spatial resolution, enabling us to map the structure and dynamics of regions that are otherwise unresolvable.

However, several aspects of the innermost AGN structure, particularly in the immediate vicinity of the central SMBH, remain poorly understood. In this work, we therefore focus specifically on these central regions, with an emphasis on the location and geometry of the hot corona. To avoid complications introduced by obscuration or jet-dominated emission, we restrict our analysis to Type 1 AGNs, in which the nuclear components are most clearly visible.

The physical nature and geometry of the hot corona have been debated since its earliest formulation \citep{haardt1991,Haardt1993}. The initial plane-parallel configuration proposed in these models was unable to reproduce the extremely hard X-ray power-law spectra observed in some sources, prompting the development of more general geometrical prescriptions \citep[e.g.][]{svensson1984,Haardt1994}. Subsequent models incorporated increasingly sophisticated treatments of Comptonization and pair-production cooling \citep[e.g.,][]{poutanen1996}, significantly improving the physical realism of corona simulations. Nonetheless, the stationary spectral shape alone has proven insufficient to uniquely determine the true geometry of the corona, leaving key aspects of its structure unresolved. Consequently, most studies adopt the lamp-post model as a practical approximation, despite the fact that several works have highlighted significant limitations and potential inconsistencies associated with this approach \citep{Zoghbi2021, 2023MNRAS.525.4524G}.

New insights into AGN inner structure have emerged from spectral studies of disk irradiation by the corona, particularly those incorporating relativistic effects and modeling the formation of the relativistically smeared Fe K$\alpha$ line \citep{george1991,campana1995,dovciak2004}. For such studies, a simplified description of the corona as a compact, point-like X-ray source located on the rotational axis, the so-called lamp-post model, has generally provided an adequate first-order approximation \citep[][]{Matt1991, Martocchia_Matt_1996}. Using this framework, numerous works have derived corona heights and reflection signatures for a wide range of AGNs \citep[][]{george1991, Matt1991, Martocchia_Matt_1996, Reynolds1997, Miniutti_Fabian_2004, 2013MNRAS.430.1694D, 2014MNRAS.439.3931E, kammoun2019, ursini2020}. However, in some cases the basic lamp-post formulation required further refinement, either by modifying the disk structure (e.g. invoking an inner hot flow, \citealt{lohfink2013,fabian2014,serafinelli2023}) or by allowing the corona itself to be dynamic (outflowing or inflowing, \citealt{Beloborodov1999}). Such modifications help to resolve discrepancies between the predicted and observed normalization of the reflected component relative to the incident continuum, an issue that can be naturally explained by Doppler boosting of radiation away from or toward the disk \citep{Beloborodov1999}.

Building upon these studies, long X-ray monitoring campaigns enabled the measurement of both Fourier-resolved time delays and energy-dependent time delays \citep[see][for an overview]{uttley_rev_2014}, a technique hereafter referred to as X-ray$-$RM. The first hard lags were detected in Cygnus X-1 by \citet{Miyamoto} and subsequently in many other AGNs \citep[e.g.,][]{papadakis2001, 2004MNRAS.348..783M}, but they seemed to be related rather to spectral evolution within the comptonizing medium \citep{2019MNRAS.487..667C, 2023MNRAS.519.4951Z}. Later, however, soft X-ray lags were discovered \citep{Fabian2009}, which are consistent with reprocessing in the inner accretion disk \citep{2011MNRAS.412...59Z}. Importantly, these soft lags were found to scale approximately linearly with the black hole mass ($M_{\text{BH}}$) and can be interpreted in terms of the corona height ($h$) above the disk \citep[e.g.,][]{deMarco2013}. The measured lag amplitudes range from roughly 100 s to 500 s \citep{kara2016}, depending on $M_{\text{BH}}$, formally corresponding to light-travel distances of less than $\sim$ 6 $r_g$. Lags associated with the Fe K$\alpha$ line were also measured, indicating slightly larger delays, typically in the range of 1$-$10 $r_g$ \citep{zoghbi2013, kara2013}. Comparison of observational results with detailed modeling has, however, remained relatively uncommon. \citet{caballero2020} successfully fitted both the lag and energy spectra of IRAS~13224-3809, demonstrating that consistent modeling is possible in at least some cases. In contrast, \citet{caballero2018} reported limitations of the simple lamp-post geometry for three additional sources, and \citet{Zoghbi2021} likewise found substantial difficulties with the lamp-post model in the two systems they analyzed. Specifically, MCG-5-23-16 showed no evidence of relativistic reverberation, while SWIFT~J2127.4$+$5654 exhibited some signatures of such behavior, but only for parameter values inconsistent with the time-averaged spectrum. The delay of the K$\alpha$ line was reproduced within a lamp-post framework, but only over a restricted frequency range, and the inferred corona height was below 10 $r_g$ \citep{epitropakis2016}.

Most of these results were derived assuming a compact, lamp-post-like corona, although extended corona models have occasionally been explored in this context \citep{Hancock2023}. These studies demonstrated that both the height and radial extension of the corona can vary over time, as illustrated in the long-term monitoring of 1H~0707$-$495 and IRAS~13224$-$3809.

Independent of the X-ray studies, the lamp-post model has also been widely applied to explain UV-optical variability as the result of disk irradiation by a variable X-ray source \citep[e.g.,][]{rokaki1993}. This framework quickly became a standard approach for modeling UV-optical inter-band time delays measured in photometric continuum$-$RM campaigns \citep[continuum$-$RM;][]{2005ApJ...622..129S, 2014MNRAS.444.1469M, shappee2014, fausnaugh2016, 2018ApJ...862..123M, 2021ApJ...922..151K, 2022ApJ...940...20G,  edelson2024, 2025ApJ...985...30M,  2025A&A...700L...8P, 2025arXiv251213296M}, particularly after the inclusion of the corona height as a key parameter \citep[e.g.,][]{cackett2007,starkey2017,kammoun2019,kammoun2021,kammoun_analit2021}.

A natural question arises: are the constraints on the corona height derived from X-ray$-$RM studies and those obtained from UV–optical time delay fitting in photometric continuum$-$RM consistent for a given source? At present, there is no clear answer. Historically, these two lines of investigation, i.e., corona height measurements from X-ray$-$RM and from UV–optical continuum$-$RM have been conducted independently, often for different sources and using distinct models or software packages. Consequently, the methodologies employed in X-ray$-$RM and UV–optical continuum$-$RM differ substantially in their fitting procedures and analysis frameworks. To achieve a consistent physical interpretation, it is essential to revisit these approaches and analyze them using a unified methodology.

In the present work, we adopt a consistent modeling framework for both spectral and timing analyses across the X-ray and UV–optical spectral ranges. The model was originally developed by \citet{dovciak2004} for modeling X-ray spectra with full relativistic effects and includes an option for time-resolved spectral modeling. This framework was later extended to UV–optical time delay fitting by \citet{kammoun_analit2021} and \citet{dovciak_KYNSED_2022}. Our goal is to use this uniform framework to test whether a single set of global parameters, such as the black hole mass, accretion rate, and corona height can simultaneously reproduce the observed properties inferred independently from X-ray$-$RM and UV-optical continuum$-$RM using both spectral and time-domain data. Here, we present the first results obtained for one of the most intensively studied sources, the Seyfert 1 galaxy NGC~5548. The paper is organized as follows. Section~\ref{ss:mod} describes the model employed to fit both the X-ray and UV–optical spectral and time-domain data. Section~\ref{ss:samp} provides details of the selected AGN and the observational data used in the analysis. The analysis and the resulting findings are presented in Section~\ref{ss:result}, followed by a discussion in Section~\ref{ss:dis}. Finally, the main findings are summarized in Section~\ref{sec:sum}.

\section{Model}
\label{ss:mod}

For our analysis, we employ the publicly available code {\tt KYNXiltr}\footnote{\url{https://projects.asu.cas.cz/dovciak/kynxiltr}}, described in \citet{kammoun_code_2023}. This code builds upon {\tt KYNSED}, the spectral model introduced by \citet{dovciak_KYNSED_2022}, with several components first presented in \citet{kammoun_analit2021}. The {\tt KYNXiltr} framework has been used to model UV–optical inter-band delays in nine AGNs, including NGC~5548 \citep{kammoun2021, kammoun_code_2023}. Additionally, \citet{2024A&A...691A.252L} applied the same approach to fit the time lag measurements of luminous AGNs and found that a corona height of roughly more than $40~r_g$ can reasonably account for the observed delays. As the code has undergone continuous development over the years, we summarize below the key assumptions underlying the current version relevant to our work.

The model fully incorporates General Relativity (GR) effects and adopts a specific geometric configuration. It assumes a lamp-post geometry for the hot corona, making the corona height an explicit free parameter. Disk irradiation is calculated under the assumption of a point-like source. However, the corona cannot be accurately described as point-like, its finite size plays a crucial role in shaping the hard X-ray spectrum. In particular, the size of the corona determines the fraction of soft photons from the disk that are Comptonized within it. This effect is explicitly accounted for in the code. The hard X-ray emission from the corona is characterized by its spectral slope and high-energy cut-off, while the low-energy cut-off is determined by the temperature of the seed photons originating from the accretion disk. Because the disk temperature varies with radius, the disk emission is radially integrated to obtain the overall spectral shape and, consequently, the effective seed-photon temperature for Comptonization. The corona is assumed to emit isotropically in its rest-frame.

The energy emitted by the hot corona constitutes a fraction of the total accretion energy budget. This energy is supplied by a Keplerian accretion disk of negligible geometrical thickness. Consequently, the global physical parameters of the model, such as $M_{\text{BH}}$, accretion rate ($\dot{m}$ in units of Eddington ratio), black hole spin ($a*$), and the fractional division of accretion power between the corona and the disk ($L_{\text{transf}}/L_{\text{disk}}$), jointly determine the overall solution.

 In principle, the model also allows for an alternative scenario in which the corona is powered by a mechanism independent of the accretion disk. Such power could be extracted from the black hole spin or arise from processes operating within the innermost stable circular orbit (ISCO). However, we do not consider this alternative configuration in the present analysis.

In the lamp-post geometry, the compact X-ray source illuminates the accretion disk, where the incident radiation is partly reflected and partly absorbed and re-emitted, thereby modifying the local disk temperature. The disk albedo varies with radius and is computed under the assumption of a constant number density in the disk atmosphere, while accounting for the appropriate local flux. The underlying atomic physics is taken from the {\tt XillverD } tables of \citet{garcia2016}, calculated for a representative density of $n = 10^{15}$ cm$^{-3}$. The ionization state of the disk is determined by the local (radial) dependence of the reprocessing efficiency on the incident flux, which naturally leads to an ionization parameter that varies with radius.

The code tracks photon trajectories, allowing it to follow not only spectral modifications but also the associated time delays. This framework enables the computation of the broadband spectral energy distribution (SED) using {\tt KYNSED}, as well as the time dependent response of disk to an impulsive flash from the corona using {\tt KYNXiltr}. The latter provides the wavelength-dependent response function of the disk. Additionally, geometrical effects, such as the viewing angle of an observer, are fully incorporated into the calculation.

The lag–spectrum modeling is carried out using two complementary approaches: one designed for X-ray$-$RM, where energy and frequency- resolved lags are measured, and another for modeling UV/optical inter-band time delays. This separation reflects the distinct observational strategies required in these two wavelength regimes. Nonetheless, a single physical model (packages inside {\tt KYNXiltr} and {\tt KYNSED}) is employed to ensure consistency and to minimize potential systematics arising from heterogeneous analysis methods.

In the X-ray band, we perform a Fourier analysis of the model transfer functions. This procedure provides Fourier-resolved time delays within selected energy intervals, as well as energy-resolved delays computed over specific frequency ranges. These quantities are directly comparable to those obtained from the observational data.

In the UV-optical bands, the mean arrival time of the signal at each wavelength is computed directly from the corresponding transfer function. Inter-band time delays are then obtained by measuring these mean arrival times relative to a chosen reference band. The resulting model-derived delays can be directly compared with the inter-band time delays inferred from the observational data.

\begin{table}[h]
\centering
\caption{Best-fit parameters inferred from the time-resolved broadband X-ray/UV/optical SED analysis}
\label{tab:kammoun24}
\centering
\begin{tabular}{ccccc} \hline \hline

% ID & & & \multicolumn{2}{c}{ICCF} & \multicolumn{2}{c}{\textsc{Javelin}} \\

 Data set & $L_{\text{transf}}/L_{\text{disk}}$ & $h$ & $\Gamma$ \\
  (MJD) &  & ($r_g$) &    \\
(1) & (2) & (3) & (4) 
\\ \hline

 S1 (56712.9 $-$ 56714.2)  &  0.90 & 40.9 & 1.52 \\
 S2 (56721.8 $-$ 56723.2) & 0.81 & 3.9 & 1.71 \\
 S3 (56731.6 $-$ 56732.9) &  0.79 & 4.3 & 1.60 \\
 S4 (56740.0 $-$ 56741.3) & 0.76 & 15.8 & 2.09  \\
 S5 (56747.9 $-$ 56749.2) & 0.64 & 27.3 & 1.89  \\
 S6 (56752.7 $-$ 56754.0) &  0.88 & 44.8 & 1.26 \\
 S7 (56755.7 $-$ 56757.0) &  0.88 & 38.3 & 1.49 \\
 S8 (56758.8 $-$ 56760.0) & 0.73 & 7.3 & 1.90 \\
 S9 (56769.7 $-$ 56771.0) & 0.58 & 12.0 & 1.90 \\
 S10 (56779.2 $-$ 56780.0) & 0.70 & 20.5 & 1.60 \\
 S11 (56788.0 $-$ 56789.0) & 0.61 & 13.4 & 1.56 \\
 S12 (56798.7 $-$ 56800.0) &  0.55 & 11.5 & 1.73 \\
 S13 (56810.7 $-$ 56812.0) &  0.68 & 23.1 & 1.54 \\
 S14 (56823.4 $-$ 56824.7) & 0.68 & 48.0 & 1.88 \\
 S15 (56827.2 $-$ 56828.5) &  0.78 & 36.3 & 1.37 \\

\hline

\end{tabular}
\vspace{0.01cm}

\tablefoot{Columns are: (1) Data sets corresponding to different segments of the UV$-$X-ray light curves, (2)  the fractional
division of accretion power between the corona and the disk, (3) corona height, and (4) photon index. The values of the parameters are retrieved from \citet{Kammoun2024}.}

\end{table}

\section{Sample and Data}
\label{ss:samp}

We therefore focused on identifying sources for which both X-ray Fourier- and energy-resolved delays and UV-optical inter-band time delays had been measured. However, such objects remain rare.

\citet{kara2016} conducted X-ray$-$RM of Seyfert 1 galaxies using archival observations from the XMM-Newton observatory that were publicly available up to January 1, 2015. Their final sample included 43 Seyfert galaxies exhibiting a wide range of flux levels, exposure times, and variability strengths. They reported both frequency- and energy-resolved lags, along with the X-ray SED for this sample. To explore possible connections between X-ray and longer-wavelength reverberation signatures, we cross-matched this sample with AGNs for which continuum UV–optical time-delay measurements are available from various photometric continuum$-$RM campaigns in the literature. This comparison yielded three common sources: NGC~5548, NGC~4151, and Mrk~335. 

Among these, NGC~5548 stands out as having the highest-quality data for both X-ray$-$RM and UV-optical continuum delays. The latter were obtained by \cite{fausnaugh2016}, who carried out an extensive photometric monitoring campaign combining space-based HST observations from the HST/COS UV RM program with simultaneous ground-based observations from sixteen observatories. The optical data were obtained in multiple broadband filters, including Johnson/Cousins $BVRI$ and Sloan Digital Sky Survey (SDSS) $ugriz$, covering the period from December 2013 to August 2014. We therefore adopt the results of the X-ray analysis presented by \citet{kara2016}, including their source spectrum, frequency and energy resolved lag-spectra (see the corresponding panel in their Figure~A1). These results are based on a long XMM-Newton light curve with a duration of $9.55 \times 10^4$ s. We note that we did not perform an independent analysis of these data.

Furthermore, \cite{Kammoun2024} investigated the time-averaged and variable broadband X-ray/UV/optical SEDs of NGC~5548 using data from Swift, HST, and ground-based facilities. Their goal was to test whether the observed broadband spectral behavior could be explained by the X-ray illumination scenario, despite the relatively modest correlation observed between X-ray and longer-wavelength variations. For our analysis, we used their best-fit results obtained from individual spectral fits with the {\tt KYNSED} model, which are summarized in Table~\ref{tab:kammoun24}. Additionally, we utilized the broadband SED provided by \citet{mehdipour2015} in our analysis.

Given its extensive multi-wavelength coverage and well-characterized variability properties, NGC~5548 thus serves as an ideal laboratory to test the consistency between  SED fitting and lag–spectrum analysis when both are derived from a coherent, homogeneous approach.

\section{Analysis and Results}
\label{ss:result}

The number of free parameters in the adopted model is considerable, including, $M_{\text{BH}}$, $\dot{m}$, $a*$, $h$, $L_{\text{transf}}/L_{\text{disk}}$, photon index ($\Gamma$), extinction parameter determined by $\mathrm{E(B-V)_{host}}$, among others. Consequently, instead of performing a blind search for parameter combinations capable of reproducing the broadband SED, the X-ray Fourier- and energy-resolved time delays, and the UV–optical inter-band delays, we adopt a multi-step strategy.

First, we fix most of the global model parameters based on the broadband X-ray/UV/optical SED fitting of NGC~5548 presented by \citet{Kammoun2024}, who employed the same physical model. Using these parameters, we compute the X-ray properties predicted by the SED fit and compare them directly with those inferred from the observed X-ray timing analysis.

Next, we explore how the model-predicted X-ray characteristics depend on key parameters, identifying which ones primarily govern the agreement between the model and the data.

Finally, we invert the procedure: starting from the parameters that best reproduce the X-ray timing properties, we examine how this parameter set translates into UV–optical behavior and assess its consistency with the observed inter-band delays of the source.

\begin{figure}
\centering
\includegraphics[width=0.45\textwidth]{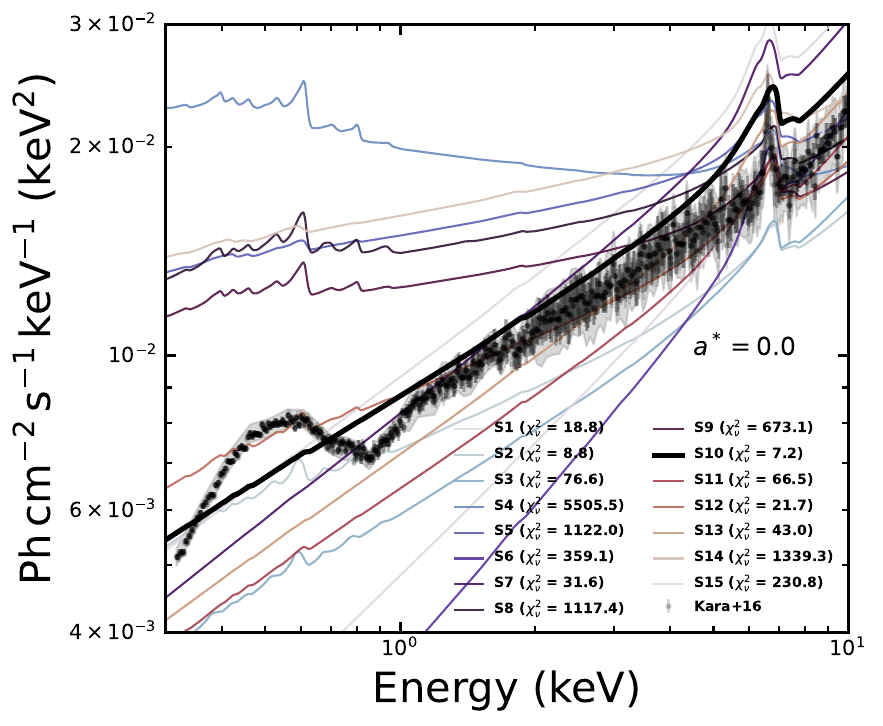}
\includegraphics[width=0.45\textwidth]{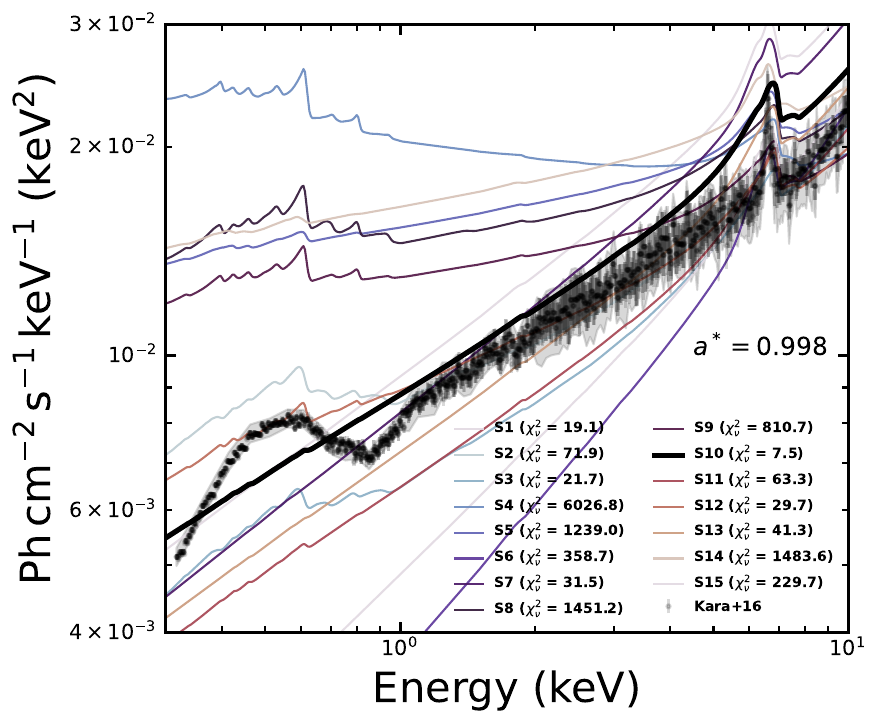}
\caption{Model fits overlaid on the observed X-ray SED for $M_{BH} = 7 \times 10^7$, $\dot{m}$=0.05, $i = 40^{\circ}$, and $f_{\text{col}} =1.7$. Top: results for spin $a*=0.0$. Bottom: results for $a*=0.998$. Representative data points from the observed X-ray SED reported in \citet{kara2016} are shown as black points, and the gray shaded region indicates their vertical range. Model curves corresponding to different parameter sets are plotted in various colors, while the best-fitting model with the lowest reduced $\chi^2_{\nu}$ is shown in black.}
 \label{fig:xray_spc}
\end{figure}

\begin{figure}

\centering
\includegraphics[width=0.45\textwidth]{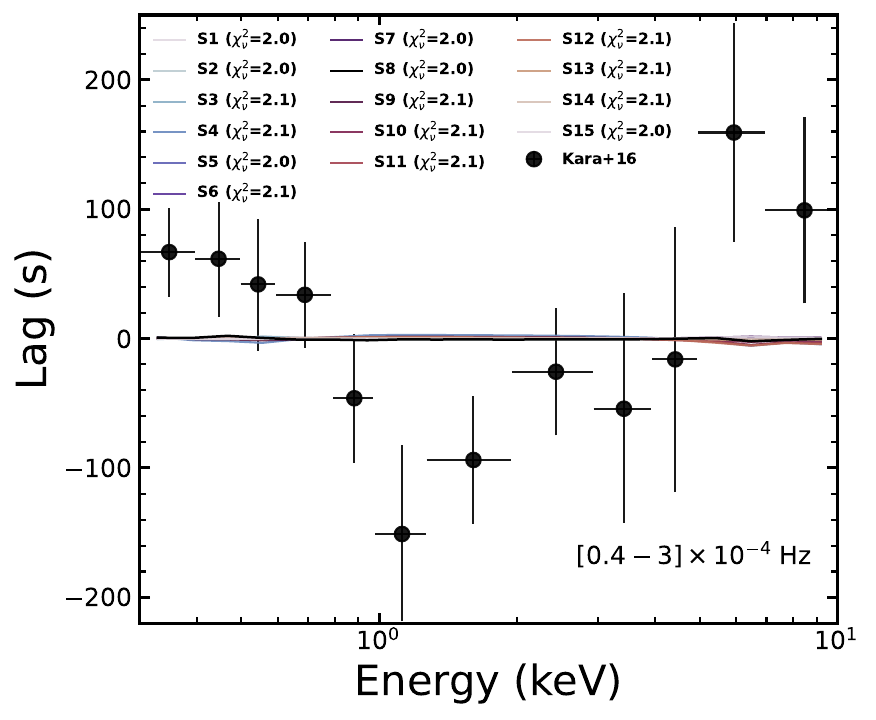}
\includegraphics[width=0.45\textwidth]{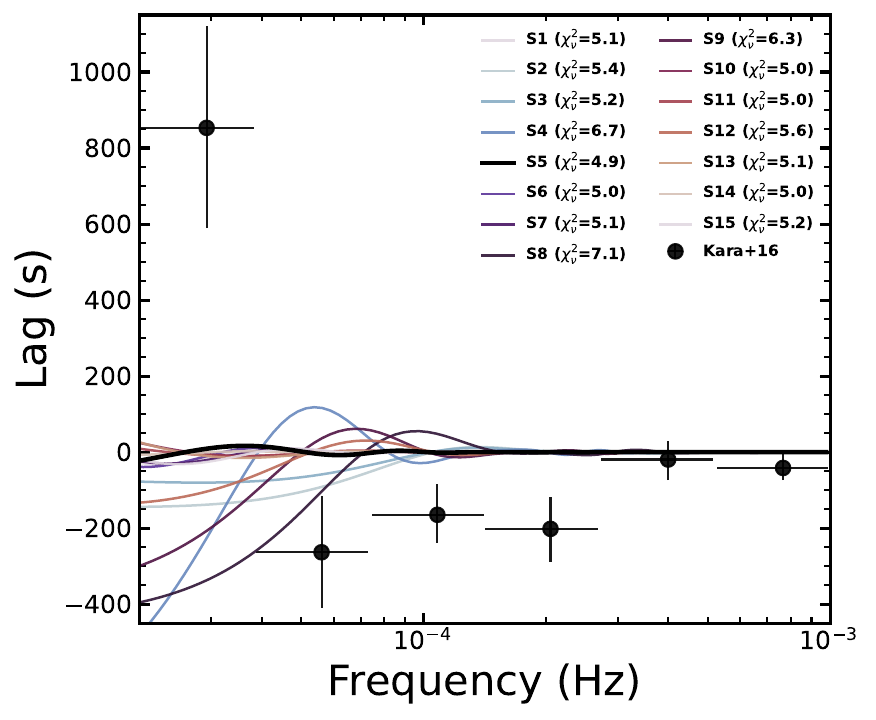}

\caption{Top: model-fitted lag–energy spectra overlaid on the observed values (black points with error bars) from \citet{kara2016} for $M_{BH} = 7 \times 10^7$, $\dot{m}$=0.05, $i = 40^{\circ}$, $a*=0.0$, and $f_{\text{col}} =1.7$. Bottom: model-fitted lag–frequency spectra in different colors for the same set of input parameters, overlaid on the observed data shown as black points with error bars. Model curves for different parameter sets are shown in various colors, with the best-fitting model corresponding to the lowest reduced $\chi^2_{\nu}$, highlighted in black.}
\label{fig:xray_rm}
\end{figure}

\subsection{Testing X-ray properties at the basis of the broad-band SED fitting}
\label{ss:xraysed}

We extracted several representative data points from the observed X-ray SED of NGC~5548 presented in \cite{kara2016}. These are shown as black points in Figure~\ref{fig:xray_spc}, while the gray-shaded region marks the vertical extent of the original data. Since \cite{kara2016} did not provide the underlying numerical values, we digitized the spectrum directly from their published figure. Using these extracted points, we calculated the reduced $\chi^2$ per degree of freedom ($\chi^2_\nu$) and the corresponding upper-tail $\chi^2$ probability ($p$-value) to enable a rough comparison with our model fits. This approach provides only an approximate assessment; a more accurate and robust comparison would require re-deriving the original data from \cite{kara2016} using a consistent reduction and calibration procedure, which lies beyond the scope of the present work. Nevertheless, the constructed spectrum clearly shows two prominent features: a soft X-ray excess below approximately 1 keV and a broad Fe K$\alpha$ emission line centered near 6.4 keV.

Building on this, we calculated the model predictions as based on the optical best fitting results.
We adopted the same set of input parameters as in \cite{Kammoun2024}, generating 15 different models corresponding to 15 segments of the UV$-$X-ray light-curve data studied independently by \citet{Kammoun2024}. Those segments were characterized by various combinations of the corona height $h$, photon index $\Gamma$, and energy fraction going to the hot corona and referred to as transfer ratio $L_{\text{transf}}/L_{\text{disk}}$, as they evolved with time (see Table~\ref{tab:kammoun24} for reference). The $M_{\text{BH}}$ was fixed at $M_{\text{BH}} = 7 \times 10^7 \, M_{\odot}$, with an accretion rate of $\dot{m} = 0.05$ and an inclination angle of $i = 40^{\circ}$, consistent with the values adopted by \cite{Kammoun2024}. We adopted a constant color-correction factor of $f_{\text{col}} =1.7$ \citep{1995ApJ...445..780S, 2001MNRAS.325.1266Z, 2005ApJ...625..913H, kammoun_code_2023, Kammoun2024} throughout our analysis.

We first show the resulting X-ray spectral models for different parameter sets in Figure~\ref{fig:xray_spc}, with colors representing distinct combinations of input parameters. The top and bottom panels correspond to spin values of $a* = 0$ and $a* = 0.998$, respectively. For each model, we calculated the reduced $\chi^2_{\nu}$ with $p$-value to evaluate the fit quality. The best-fitting model, corresponding to the lowest $\chi^2_{\nu}$, is highlighted with a solid black line.

For both the non-spinning ($a* = 0$) and maximally spinning ($a* = 0.998$) cases, we obtained    largely consistent results, with a few exceptions reflected in the significant differences in the corresponding reduced $\chi^2_{\nu}$ values, and in both cases the best fit is provided by parameter set S10, characterized by $h = 20.5 \, r_g$, $\Gamma = 1.60$, and $L_{\text{transf}}/L_{\text{disk}} = 0.70$. 

However, for the parameter sets considered, the models systematically deviate from the observed spectrum, resulting in $p$-values close to zero. In particular, they fail to reproduce the soft X-ray excess, and the best-fitting model tends to overestimate the flux at higher energies, indicating that additional physical components or alternative parameter configurations may be needed to accurately describe the X-ray emission of NGC~5548. Moreover, the modeled Fe K$\alpha$ line appears broader than observed. Although the X-ray and optical data are not from the same epoch, variability alone is unlikely to account for the discrepancies between the model and the observations. However, \citet{Kammoun2024} accounted for intrinsic X-ray absorption, including possible warm absorbers while fitting the broadband SED, whereas our model fitting does not incorporate any intrinsic absorption component. %It also remains unclear whether \citet{kara2016} accounted for intrinsic absorption when constructing their X-ray spectrum. 
\citet{kara2016} did not explicitly model the X-ray flux spectrum, but rather presented the observed X-ray spectral data. Consequently, absorption effects are not accounted for in the X-ray spectral data used in our study. Therefore, inconsistencies already exist between the data and the parameter sets adopted in  modeling the spectrum, which may contribute to the discrepancies observed in our comparisons.

 To assess whether the soft X-ray excess and the potential X-ray obscuration feature around 0.9 keV influence the comparison between our model predictions and the observed X-ray spectrum, we performed an additional test in which the spectral fitting was restricted to the 1.5$-$10 keV energy range. %We find that excluding the lower-energy band does not affect the main conclusions derived from the fits over the full energy range. 
The only notable change is a systematic reduction in the reduced $\chi_{\nu}^{2}$ values across all parameter sets, resulting in a better fit to the data. In the restricted-band analysis, parameter set S12 (S2) provides the best fit, yielding $\chi_{\nu}^{2}=0.6$ ($0.2$) for $a* = 0$ ($a* = 0.998$), whereas parameter set S10 was preferred when the full 0.3–10 keV range was considered.

Nevertheless, our primary objective is to reproduce the observational framework of \cite{kara2016}, which includes both energy- and frequency-resolved lag-spectra derived from the entire 0.3$-$10 keV band. Restricting the analysis to the 1.5$-$10 keV range would therefore prevent a comprehensive comparison between the model predictions and the full set of observational constraints provided by the X-ray lag-spectra. Consequently, we do not pursue the restricted-band analysis further and instead focus on results obtained over the complete energy range.

\begin{figure}[h!]
\centering
\includegraphics[width=0.45\textwidth]{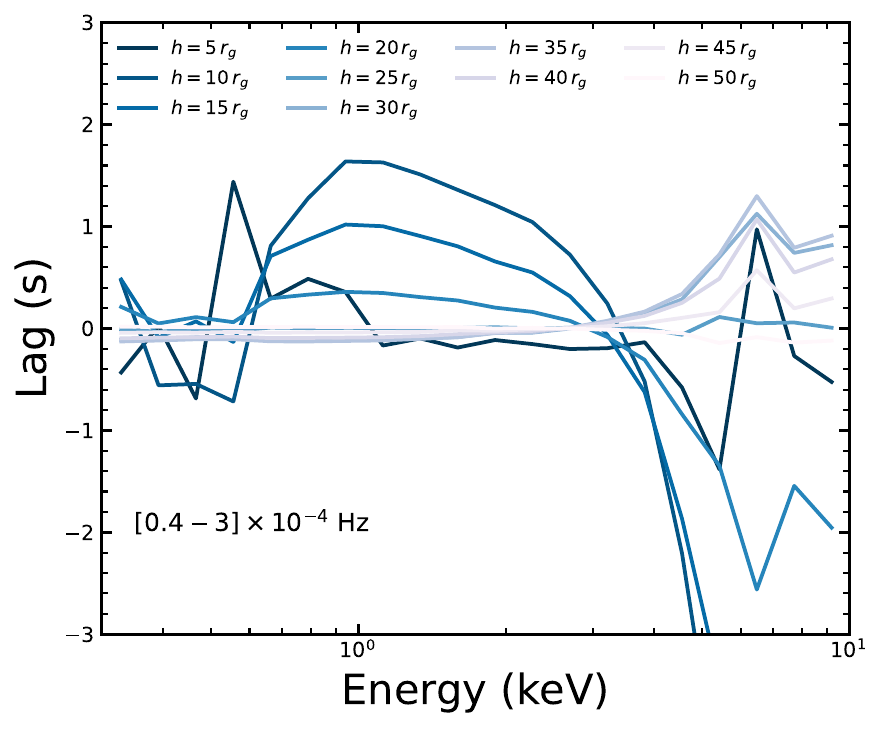}
\includegraphics[width=0.45\textwidth]{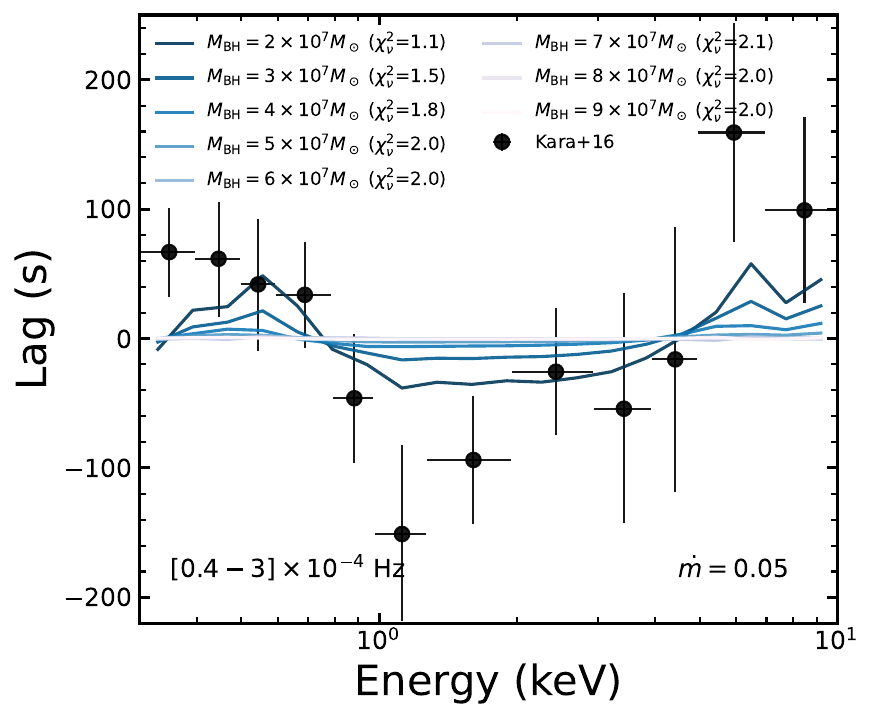}
\includegraphics[width=0.45\textwidth]{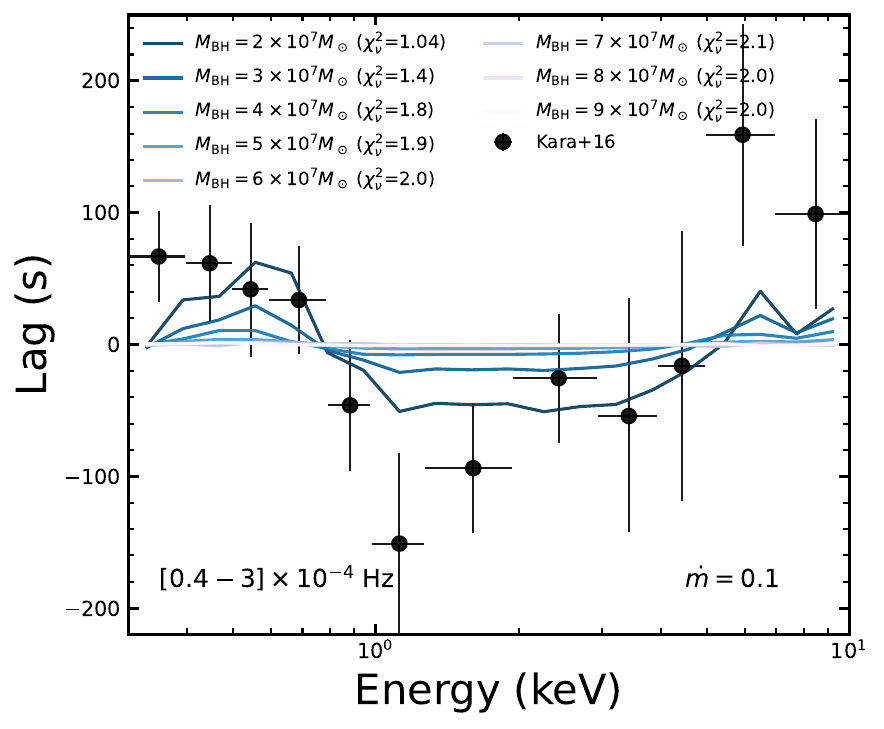}

\caption{Top: model-fitted lag–energy spectra for different corona heights $h$, assuming fixed parameters of $M_{\mathrm{BH}} = 7 \times 10^7 \, M_\odot$, $\dot{m} = 0.05$, $i = 40^{\circ}$, $\Gamma = 1.58$, $L_{\mathrm{transf}}/L_{\mathrm{disk}} = 0.9$, and $f_{\text{col}} =1.7$. Middle: model-fitted lag–energy spectra for different $M_{\mathrm{BH}}$, with $\dot{m}=0.05$, a constant  height of $h = 5 \, r_g$ and other parameters fixed to the above values. Bottom: same as middle panel but for $\dot{m}=0.1$. The observed data points are shown by black points with error bars in the middle and bottom panel.}
\label{fig:model_fit}
\end{figure}

Next, we employed the code to model the lag–energy and lag–frequency spectra of NGC~5548 obtained from X-ray$-$RM. In the lag–energy spectrum, time lags are computed over a specific frequency range  ([$0.4 - 3] \times 10^{-4}$) Hz between a narrow band of interest ($0.3-1$ keV) and a broad reference band covering the full $0.3-10$ keV range. Similarly, the lag–frequency spectrum is derived by measuring the lags between the soft ($0.3-1$ keV) and hard ($1-4$ keV) energy bands. These energy ranges were chosen to maintain consistency with those adopted by \cite{kara2016}. During the fitting, we used the same set of input parameters described above.

The resulting model fits are shown in Figure~\ref{fig:xray_rm}, with the top and bottom panels corresponding to the lag–energy and lag–frequency spectra, respectively. We find that the model fails to reproduce the observed data in both spectra when the input parameters from Table~\ref{tab:kammoun24} are adopted. In the lag–energy spectrum, the model remains nearly constant across the entire energy range and does not capture the negative lags observed between $\sim 0.8$ and 5 keV.  For example, parameter set S8 yields the lowest $\chi^2_{\nu} = 2.0$ corresponding to a $p$-value of 0.022. Similarly, in the lag–frequency spectrum, the model cannot reproduce the observed negative lags between $\sim 5 \times 10^{-5}$ and $10^{-3}$ Hz, while the model predicts negative time lags at much lower frequencies, resulting in $p$-values less than 0.001.

In summary, these results indicate that the current parameter sets are insufficient to fully capture the observed X-ray$-$RM signatures of NGC~5548. This highlights the need for additional physical components, alternative parameter configurations, or a more complex treatment of the corona and disk structure to accurately model the lag–energy and lag–frequency behavior.

\subsection{X-ray lag-energy and lag-frequency dependence on the global parameters}
\label{sect:various}

To investigate the source of the discrepancy, we relax the model constraints imposed by \citet{Kammoun2024} and listed in Table~\ref{tab:kammoun24}, and focus on identifying the key parameters that drive the predicted X-ray behavior. The model depends on several parameters, including $M_{\text{BH}}$, $\dot{m}$, $h$, $i$, $L_{\text{transf}}/L_{\text{disk}}$, and $\Gamma$. Among these, $M_{\text{BH}}$, $h$, and $\dot{m}$ are expected to have the strongest influence on the model predictions. To assess their impact, we first examine each parameter individually, beginning with the corona height.

To study the dependence of the lag–energy spectrum on the corona height $h$, we generated model spectra for a range of $h$ values from 5 $r_g$ to 50 $r_g$, keeping all other parameters fixed at their previous values (e.g., $M_{\text{BH}} = 7\times10^7 \, M_\odot$, $\dot{m}=0.05$, and $i = 40^{\circ}$). We also fixed the photon index to $\Gamma = 1.58$, consistent with the observed value of $\Gamma = 1.58 \pm 0.02$ reported for NGC~5548 by \citet{2015A&A...577A..38U}, and adopted $L_{\text{transf}}/L_{\text{disk}} = 0.9$. The resulting models are shown in the top panel of Figure~\ref{fig:model_fit}. As the corona height increases, the model-predicted lag amplitude near 1 keV becomes progressively larger and changes from negative to positive values, deviating further from  the observed lag–energy spectrum in Figure~\ref{fig:xray_rm}, upper panel. Even at the smallest height ($h=5 \, r_g$), the model still fails to reproduce the observed negative lag near 1 keV.

However, it is worth noting that the corona heights reported in the literature for NGC~5548 are generally higher. For instance, \citet{starkey2017} arbitrarily fixed the height at $6 \, r_g$, while \citet{kammoun2021} found $h \sim 29.1$ [$23.2, 58.5$] $r_g$ from UV–optical lag-spectrum fitting assuming a spin of $a*=0.0$. \citet{jaiswal2025} reported an even larger value of $h = 48.3 \, r_g$ from simultaneous UV–optical lag-spectrum and SED fitting. On the other hand, \citet{gardner2017} used $h = 10 \, r_g$, although in that case it represented a ring of matter located further out rather than a classical lamp-post corona.

Another way to increase the negative lag amplitude between $\sim 0.8$ and 5 keV, as observed in the X-ray$-$RM data, is by adjusting the black hole mass, $M_{\text{BH}}$. However, the plausible range for $M_{\text{BH}}$ cannot be arbitrarily large. $M_{\text{BH}}$ itself cannot change, but its determination highly depends on the BLR state.  Multiple RM campaigns have revealed that, in many AGNs, the measured H$\beta$ time delay can vary substantially between observing epochs, often reflecting different luminosity states. Such variations imply that BLR can undergo dynamic structural changes on timescales of just a few years \citep[e.g.,][]{obscurer2014, 2018ApJ...856..108P}. These changes could be driven by variations in the accretion rate \citep{cackett2006, 2014MNRAS.438.3340E}, inhomogeneities in the BLR gas distribution \citep{1995A&A...296..332W}, or the effects of radiation pressure \citep{2023MNRAS.520.1807C}. Nonetheless, the underlying cause of this variability is still uncertain, since the observed changes in luminosity are frequently small relative to the associated measurement errors.

NGC~5548 provides a notable example of such dynamic BLR behavior, being one of the best-studied AGNs with over twenty years of RM observations. For this source, the H$\beta$ time delay varies from 2.3 days at $L_{5100} \sim 10^{42.7} \, \mathrm{erg \, s^{-1}}$ \citep{2010ApJ...721..715D} to 21.5 days at $L_{5100} \sim 10^{43.4} \, \mathrm{erg \, s^{-1}}$ \citep{peterson2002, peterson2004}, providing a clear illustration of the well-known BLR 'breathing' phenomenon. Assuming a constant virial factor across campaigns, these variations correspond to RM-based $M_{\text{BH}}$  estimates ranging from $0.7$ to $9\times10^7 \, M_\odot$, implying a mass uncertainty of roughly one dex for NGC~5548. Note that the virial factor used to convert the virial product ($R_{\text{BLR}} v^2/G$) into $M_{\text{BH}}$ depends on the geometry and kinematics of the BLR. Consequently, it can vary across different RM campaigns, introducing an additional uncertainty of more than 0.1 dex \citep{2015ApJ...801...38W}, when a constant virial factor is assumed.

Motivated by this observed variability, we performed model fitting for different black hole masses between $2$ and $9\times 10^7 \, M_{\odot}$, keeping $h=5 \, r_g$, $\dot{m}=0.05$, $i = 40^{\circ}$, $\Gamma=1.58$, and $L_{\text{transf}}/L_{\text{disk}}=0.9$. The results, shown in the middle panel of Figure~\ref{fig:model_fit}, reveal that decreasing the black hole mass enhances the negative lag amplitudes between $\sim 0.8$ and 5 keV and provides a better match to the observed lag–energy spectrum (also see Section~\ref{ss:ap_A} and Figure~\ref{fig:ener_mas_exp} for further discussion). Consequently, the X-ray$-$RM data appear to favor a smaller black hole mass, around $M_{\text{BH}} \sim 2\times10^7 \, M_{\odot}$, yielding a reduced $\chi^{2}_{\nu} = 1.1$ and a  $p$-value of 0.362. In contrast, adopting a higher mass of $7\times10^7 \, M_\odot$, as assumed by \citet{Kammoun2024} and consistent with the value reported by \citet{horne2021} $-$ results in a poorer fit, with $\chi^2_{\nu}=2.1$ and a $p$-value of 0.020.

\begin{figure}
\centering
\includegraphics[width=0.45\textwidth]{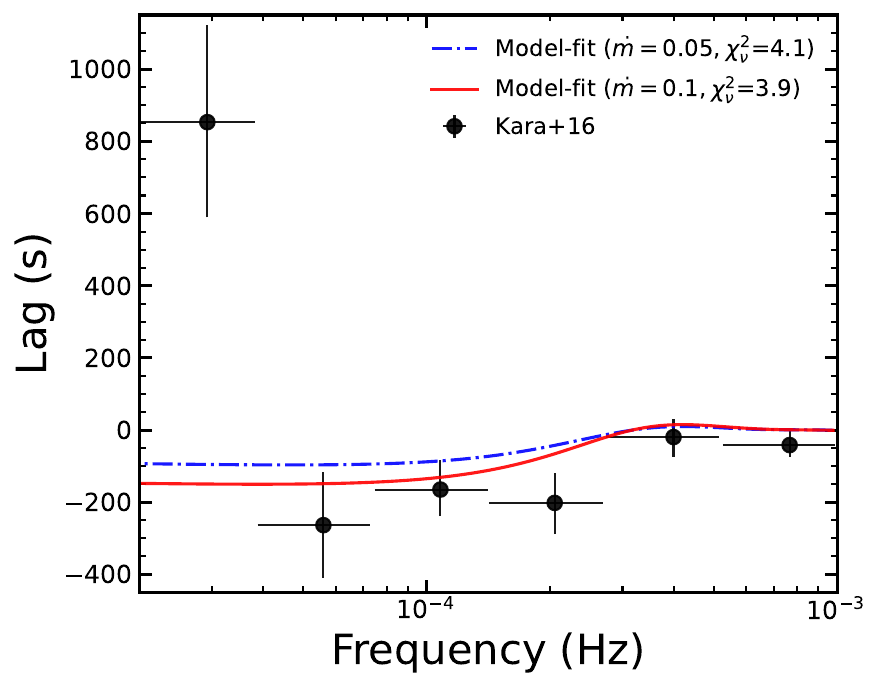}
\includegraphics[width=0.45\textwidth]{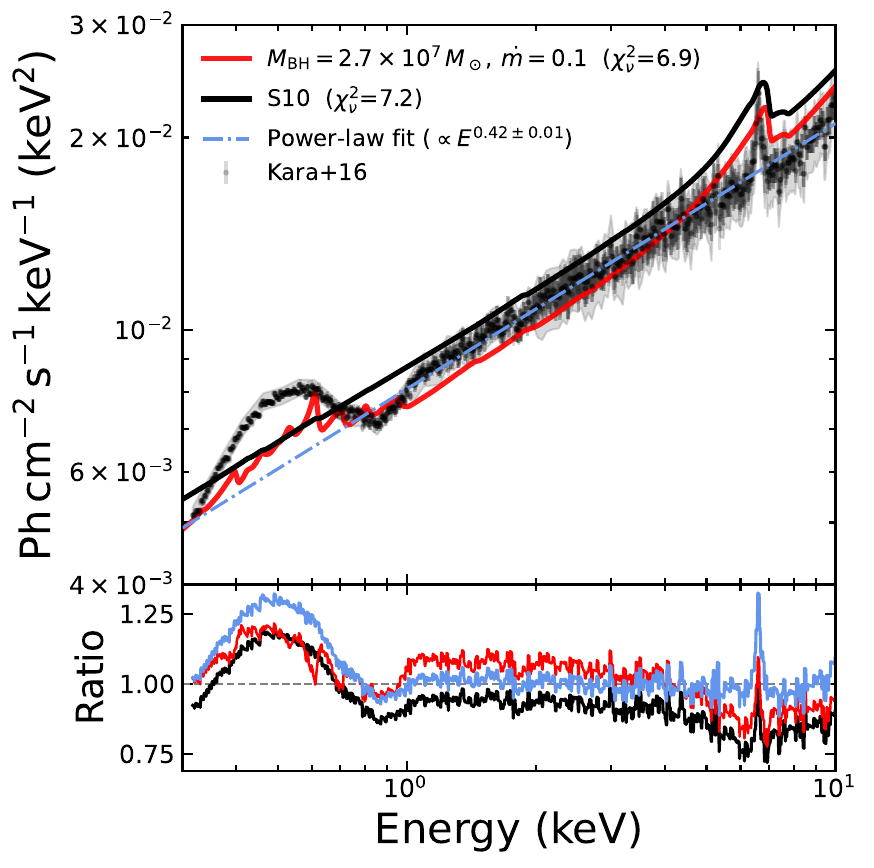}
\caption{Top: model-fitted lag-frequency spectra with $M_{\text{BH}} = 2\times10^7 \, M_\odot$, $h= 5 \, r_g$,  $i = 40^{\circ}$, $\Gamma=1.58$, $L_{\text{transf}}/L_{\text{disk}}=0.9$, and $f_{\text{col}} =1.7$ for $\dot{m}$=0.05 (blue) and 0.1 (red). Bottom: model fit on top of observed X-ray SED for $M_{BH} = 2.7 \times 10^7$, $h = 10 \, r_g$, $i=40^{\circ}$, $\dot{m}$=0.1,  $\Gamma=1.58$, $L_{\text{transf}}/L_{\text{disk}}=0.9$, and $f_{\text{col}} =1.7$  shown by red line, while for   $M_{\text{BH}} = 7 \times 10^7$,  $\dot{m}$=0.05, $h = 3.9 \, r_g$,  $i = 40^{\circ}$, $\Gamma=1.71$, $L_{\text{transf}}/L_{\text{disk}}=0.81$, and $f_{\text{col}} =1.7$ is shown by black line. The power-law best-fit is shown as a blue dot–dashed line. The lower sub-panel displays the data to model ratio for each best-fit model in their respective colors, with a horizontal dashed line indicating a ratio of unity.}
 \label{fig:xray_spc_small_mass}
\end{figure}

Further we repeated the same experiments for different values of $M_{\text{BH}}$ but with a higher accretion rate of $\dot{m}=0.1$ and the resulting model fits are shown in the bottom panel of Figure~\ref{fig:model_fit}. This time we obtain an even better model-fit for the lowest mass $M_{\text{BH}} \sim 2\times10^7 \, M_{\odot}$ with $\chi^2_{\nu}=1.04$, and a $p$-value of 0.440. Overall, the model-fits to the X-ray$-$RM data consistently favor a lower black hole mass and higher accretion rate than those inferred from broadband SED fitting.

Building upon these results, we then examined the lag-frequency spectrum using the best-fit parameters inferred from the lag-energy modeling, namely, $M_{\text{BH}} = 2\times10^7 \, M_{\odot}$, $h = 5 \, r_g$, $i = 40^{\circ}$, $\Gamma=1.58$, and $L_{\text{transf}}/L_{\text{disk}}=0.9$. To assess the effect of accretion rate on the model performance, we computed fits for both $\dot{m}$ values of 0.05 and 0.1. The resulting fits, shown in the top panel of Figure~\ref{fig:xray_spc_small_mass}, provide a significantly improved agreement with the observed data compared to the earlier models shown in the bottom panel of Figure~\ref{fig:xray_rm}. In particular, both fits reproduce the overall shape of the lag–frequency spectrum and match all data points within uncertainties, except for the large positive lag observed at the lowest frequency. Among the two cases, the model with $\dot{m} = 0.1$ yields a slightly better fit ($\chi^2_{\nu}=3.9$, $p$-value = 0.002) than that with $\dot{m} = 0.05$ ($\chi^2_{\nu}=4.1$, $p$-value =0.001), again indicating that the X-ray$-$RM data favor a smaller black hole mass accreting at a relatively higher rate.

Finally, to assess the consistency of the parameters inferred from the lag–energy and lag–frequency fits, we tested whether the same parameter set can also reproduce the observed X-ray SED. The {\tt KYNSED} model is parametrized by both $M_{\text{BH}}$ and $\dot{m}$, which primarily determine the overall normalization of the spectrum. Lowering $M_{\text{BH}}$ requires a higher accretion rate to maintain a comparable total energy flux. This adjustment, however, also introduces changes in the spectral shape. For example, at high Eddington ratios or for small $M_{\text{BH}}$, the soft X-ray excess becomes clearly visible below 1 keV, making the combined spectrum appear harder. By contrast, variations in $L_{\text{transf}}/L_{\text{disk}}$ have a much weaker impact on the spectral shape, and therefore cannot fully compensate for the changes induced by the other two parameters (see Appendix in \citealt{dovciak_KYNSED_2022}). We thus refitted the X-ray SED using parameter values close to those obtained from the timing analyses. As shown by the red solid line in the bottom panel of Figure~\ref{fig:xray_spc_small_mass}, we find that the model provides a better fit to the observed SED, with improved $\chi^2_{\nu}$ value for $M_{\text{BH}} = 2.7 \times 10^7 \, M_\odot$, $h=10 \, r_g$, $\dot{m}=0.1$, $i = 40^{\circ}$, $\Gamma=1.58$, and $L_{\text{transf}}/L_{\text{disk}}=0.9$. However, the $p$-value remains close to zero. For comparison, we also fitted the observed X-ray SED with a simple power-law model, shown by the blue dot–dashed line. The fit yields an energy dependence of $\propto E^{2-\Gamma}$, corresponding to a photon index of $\Gamma = 1.58$, which is in excellent agreement with the value reported by \citet{2015A&A...577A..38U} and adopted in our model fits. This consistency across the lag–energy, lag–frequency, and Xray-SED analyses suggests a coherent physical scenario in which NGC~5548 is better described by a relatively smaller black hole mass accreting at a comparatively higher rate.

\subsection{UV-optical lag-spectrum fitting}

In this section, we modeled the UV-optical lag-spectrum of NGC~5548 using the physically motivated X-ray reflection code {\tt KYNXiltr} \citep{kammoun_analit2021, kammoun_code_2023}, implemented within the same modeling framework used for the X-ray$-$RM data. Our goal is to test whether the physical parameters inferred from X-ray$-$RM analyses can reproduce the observed inter-band lags in the UV–optical regime. To this end, we fit the observed lag-spectrum using the same input parameters derived from our X-ray$-$RM results, as described in the previous sections. Specifically, we adopted a black hole mass of $M_{\mathrm{BH}} = 2.7 \times 10^7 \, M_{\odot}$,  $i = 40^{\circ}$, $\dot{m}=0.1$, $\Gamma = 1.58$, and  $f_{\text{col}} = 1.7$. The corona height, $h$ was treated as a free parameter, while two different ratios of, $L_{\text{transf}}/L_{\text{disk}} = 0.5$ and $0.9$, are considered.

\begin{figure}
\centering
\includegraphics[width=0.45\textwidth]{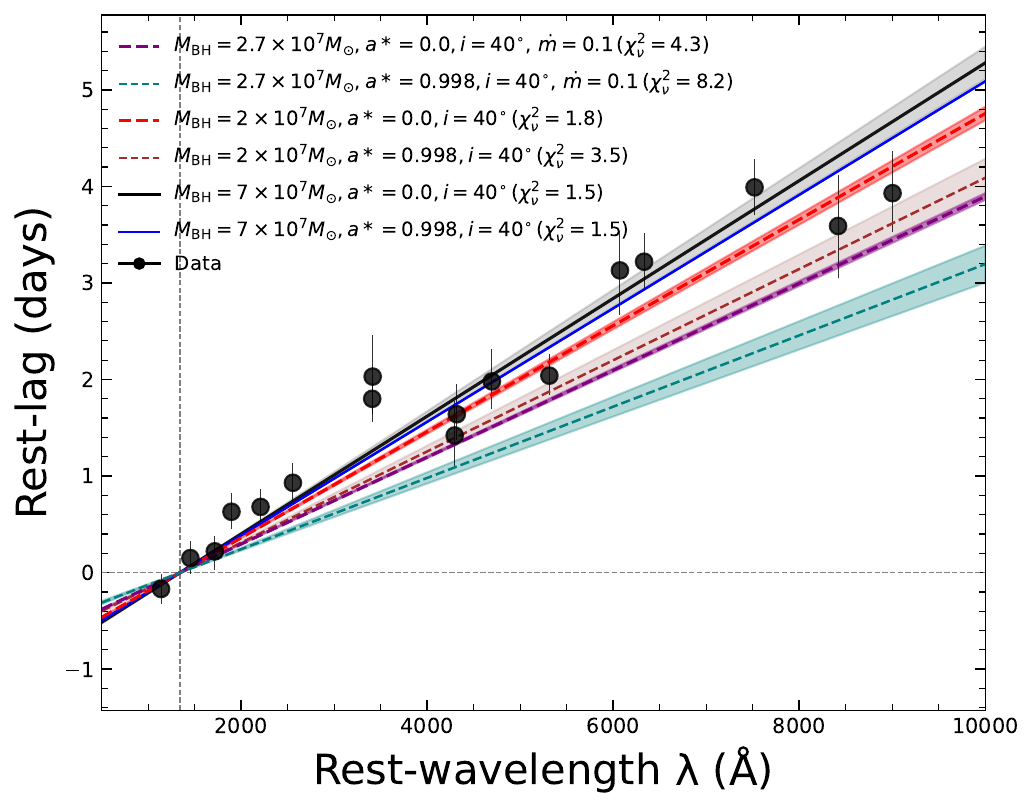}
\caption{ UV–optical lag-spectrum of NGC~5548. The black circular points with error bars show the observed rest-frame inter-band time delays reported by \cite{fausnaugh2016}. The median best-fit {\tt KYNXiltr} model curves for spins $a* = 0.0$ and $a* = 0.998$ are shown as solid and dashed lines. The fit ranges corresponding to the two adopted values of $L_{\text{transf}}/L_{\text{disk}}$ (0.5 and 0.9) are indicated by the shaded regions. Model fits for different combinations of input parameters are shown in different colors. Vertical and horizontal dotted lines indicate the rest-frame reference wavelength and zero rest-frame lag, respectively.
}
\label{fig:uv_op_lgspc}
\end{figure}

The resulting model fits for the two spin configurations, $a* = 0.0$ and $a* = 0.998$, are shown in Figure~\ref{fig:uv_op_lgspc} as dashed purple and teal lines, respectively. Although these fits employ physically consistent parameters, the model significantly underpredicts the observed UV–optical inter-band delays, resulting in reduced $\chi^2_{\nu}$ values of 4.3 and 8.2, with $p$-values less than 0.001 for the non-rotating and maximally rotating cases, respectively. Moreover, the best-fit corona heights in both scenarios exceed $90 \, r_g$, far larger than the typical values of $h \sim 5$–$10 \, r_g$ inferred from X-ray$-$RM and X-ray SED analyses. This pronounced mismatch indicates that the parameter set derived exclusively from the X-ray$-$RM data modeling is insufficient to reproduce the UV-optical lag-spectrum of NGC~5548.

To further investigate this discrepancy, we next repeated the fits using a comparable black hole mass of $M_{\mathrm{BH}} = 2 \times 10^7 \, M_{\odot}$, this time allowing both the accretion rate, $\dot{m}$ and $h$ to vary while keeping all other parameters same as before. The resulting fits, shown in Figure~\ref{fig:uv_op_lgspc} as dashed red and brown lines for $a* = 0.0$ and $a* = 0.998$, respectively, reproduce the observed inter-band delays more successfully, with improved $\chi^2_{\nu}$ values of 1.8  ($p$-value = 0.026) and 3.5 ($p$-value $< 0.001$). However, these fits require a considerably higher accretion rate, $\dot{m} \sim 0.499$, along with an increased corona height exceeding $90 \, r_g$. Despite these high values, the close match between the model and observed lag-spectrum indicates that the UV–optical delays can, in principle, be explained using a black hole mass consistent with X-ray$-$RM results, provided that both the accretion rate and corona height are significantly enhanced. However, NGC~5548 is unlikely to exhibit such a high accretion rate of 0.499 if the smaller black hole mass inferred from X-ray$-$RM is adopted.

Finally, we test whether the higher $M_{\mathrm{BH}}$ inferred from broadband SED modeling can reconcile these discrepancies. We therefore fit the observed UV–optical lag-spectrum using $M_{\mathrm{BH}} = 7 \times 10^7 \, M_{\odot}$, as used in the time-averaged and variable X-ray/UV/optical SEDs fitting of NGC~5548 by \cite{Kammoun2024}. The other parameters ($i$, $\Gamma$, $L_{\text{transf}}/L_{\text{disk}}$, and $f_{\text{col}}$) are kept the same, while $\dot{m}$ and $h$ are allowed to vary freely. With these inputs, the model once again reproduces the observed inter-band delays well as represented by the solid black and blue lines for a*=0.0, and 0.998, respectively, achieving a $\chi^2_{\nu}$ value of 1.5 with $p$-values of 0.086 and 0.090, respectively. Note that \cite{Kammoun2024} previously performed lag–spectrum fitting assuming $M_{\text{BH}} = 7 \times 10^7 \, M_{\odot}$. We repeat the same analysis here in order to ensure full internal consistency within our framework. From this fitting,  the best-fit accretion rates are $\dot{m} \sim 0.05$ for $a* = 0.0$ and $\dot{m} \sim 0.1$ for $a* = 0.998$, consistent with those inferred from broadband SED fitting by \cite{Kammoun2024}, and X-ray SED fitting only in this work, respectively. However, the corresponding corona heights remain relatively large, ranging from $\sim 23$ to $60 \, r_g$, which are still higher than the values typically inferred from X-ray$-$RM analyses, yet consistent with those reported by \citet{kammoun2021} and \citet{jaiswal2025}. The results of the UV–optical lag-spectrum model fits obtained using different values of $M_{\mathrm{BH}}$ are summarized in Table~\ref{tab:mod_uvop}.

Overall, these results reveal a systematic discrepancy between the $M_{\mathrm{BH}}$, corona heights and accretion rates derived from X-ray$-$RM and those obtained from UV–optical lag-spectrum modeling. Specifically, reproducing the UV–optical lag-spectrum requires a significantly higher and more extended corona, as well as a larger accretion rate, compared to those inferred from the X-ray domain, assuming a consistent $M_{\mathrm{BH}}$ across the two wavelength domains. In contrast, when the model fitting is performed separately within each spectral regime, i.e., X-ray (lag–energy, lag–frequency, and X-ray SED) or UV–optical (broadband SED in \cite{Kammoun2024} and UV–optical lag-spectrum), the derived parameters remain internally consistent within each respective domain. This divergence highlights the challenge of simultaneously reconciling the X-ray and UV–optical timing properties of NGC~5548 within a single, unified reprocessing framework.

\begin{table}[h!]
\centering
%\movetableright= -60mm
 \caption{ Model fitting results on UV-optical lag-spectrum and SED}
 \label{tab:mod_uvop}

\resizebox{9cm}{!}{

\fontsize{300pt}{300pt}\selectfont
\begin{tabular}{ccclr} \hline \hline

 $M_{\text{BH}} \, (M_{\odot})$ & spin & $L_{\text{transf}}/L_{\text{disk}}$ &  $h$ ($r_g$) & $\dot{m}$   
\\ 
(1) & (2) & (3) & (4) & (5) 
\\ \hline \\
& & {\tt KYNXiltr} model fit     & & \\
 & & on UV-optical lag-spectrum   & & \\ \\
 $2.7 \times 10^7$ & $a\ast=0$ & 0.5 & 97.5 & 0.1 \\
  & $a\ast=0$ & 0.9 & 97.9 & 0.1 \\
  & $a\ast=0.998$ & 0.5 & 95.5 & 0.1 \\
  & $a\ast=0.998$ & 0.9 & 90.7 & 0.1 \\  \\

 $2 \times 10^7$ & $a\ast=0$ & 0.5 & 90.8 & 0.499 \\
 & $a\ast=0$ & 0.9 & 98.1 & 0.499 \\
  & $a\ast=0.998$ & 0.5 & 99.9 & 0.499 \\
  & $a\ast=0.998$ & 0.9 & 97.7 & 0.499 \\  \\

 $7 \times 10^7$ & $a\ast=0$ & 0.5 & 23.7 & 0.054 \\
 & $a\ast=0$ & 0.9 & 58.6 & 0.063 \\
  & $a\ast=0.998$ & 0.5 & 29.9 & 0.144 \\
 & $a\ast=0.998$ & 0.9 & 21.3 & 0.131 \\ \\

\hline \\

 & &  Accretion disk + BLR reprocessing model  & & \\ 
 
&   &   fit on UV-optical lag-spectrum + SED & & \\ \\

 $2 \times 10^7$ &  &  &  34.3 & 0.053 \\

 $7 \times 10^7$ & & & 18.6 & 0.014  \\ \\

\hline
\end{tabular}
}

\tablefoot{Columns are: (1) black hole mass, (2) spin, (3) the ratio of the accretion power emitted within a specified radius to the total accretion power output,  (4) corona height in unit of gravitational radius ($r_g$), and (5) accretion rate. A constant color correction of $f_{\mathrm{col}} =1.7$, and photon index of $\Gamma=1.58$ were used during the fitting, while Accretion disk + BLR reprocessing model assumes $f_{\mathrm{col}} =1$. For accretion disk + BLR reprocessing model see Section~\ref{ss:frado}.}

\end{table}

\section{Discussion}
\label{ss:dis}

We modeled the Fourier-resolved lag–energy and lag–frequency spectra, together with the X-ray SED inferred from X-ray$-$RM and the UV–optical lag-spectrum from photometric continuum$-$RM, in a self-consistent manner using a unified set of codes within a single modeling framework. However, this approach did not produce mutually consistent results.

In particular, the Fourier-resolved X-ray delays generally imply a relatively small $M_{\text{BH}}$. However, for such a low black hole mass, the accretion rate and corona height inferred from the Fourier-resolved X-ray lag-spectra become incompatible with those derived from the UV–optical continuum$-$RM lag-spectrum in NGC~5548. The problem with the black hole mass is coupled with the problem of accretion rate.

 A second discrepancy concerns the inferred height of the corona. X-ray$-$RM studies typically favor a compact geometry, with heights of order $\sim 10 \, r_g$, as commonly reported in X-ray analyses \citep[e.g.,][]{deMarco2013}. In contrast, optical studies often imply significantly larger values, reaching up to an order of magnitude higher. While the results of \citet{kammoun_code_2023} do not always require such large heights, they nevertheless span a broad range -- from $\sim 10 \, r_g$ to $\sim 100 \, r_g$, depending on the choice of other model parameters. Most notably,  $M_{\text{BH}}$, and $\dot{m}$ should remain consistent across both X-ray and optical data sets, and the fact that they do not poses a more fundamental challenge to achieving a fully self-consistent model.

Therefore, the disagreement in the inferred black hole mass, accretion rate and corona height requires careful consideration. Several potential sources for this discrepancy include:

\begin{itemize}
\item data quality,
\item underlying assumptions of the X-ray$-$RM modeling, and
\item underlying assumptions of the UV-optical$-$RM modeling.

\end{itemize}

 We discuss the potential sources of these discrepancies below.  In this context, the present work should be regarded as a pilot study aimed at assessing the consistency of the model with the observed RM features in both the X-ray and UV–optical domains. Importantly, the lack of agreement between the model predictions and the observations is, in itself, a significant result, as it highlights the need for a more comprehensive and systematic investigation, which we intend to pursue in future work.

\subsection{Data quality}

The lag–energy and lag–frequency spectra considered in our analysis are simply taken from \citet{kara2016}, who computed them using data from a single X-ray observation. Their study relied on an XMM-Newton observation with exposure of nearly 100 ks, which is longer than typical exposures and thus offers relatively high statistical quality. Throughout the exposure, the source maintained an average count rate of about 5 counts/s. However, despite the long exposure and reasonable count rate, the reported normalized excess variance is low, $F_{\text{var}} = 0.039$. This implies that variability on timescales shorter than $\sim 150$ s corresponding to a frequency of $6.6 \times 10^{-3}$ Hz, is likely dominated by Poisson noise, although this value should be regarded only as a rough estimate. A more rigorous and comprehensive analysis of the available data is clearly required. We intend to pursue this in detail; however, the volume of data is substantial, encompassing not only observations from XMM-Newton but also datasets from multiple additional instruments. 

 For example, more robust determination of the usable frequency range requires examining the coherence function, as illustrated for example in Figure~1 of \citet{epitropakis2016}. Such tests, however, were not performed by \citet{kara2016}, leaving the reliability of the high-frequency lag measurements uncertain. Therefore, a more robust and careful analysis of X-ray data spanning a longer temporal baseline, potentially incorporating monitoring data from Swift is required. This is particularly important for NGC~5548, which hosts a relatively massive black hole and thus does not exhibit strong variability on timescales shorter than a day. We leave this detailed investigation to future work.

The optical data are of very high quality. \citet{Kammoun2024} independently analyzed several data sets and successfully modeled the SEDs. Moreover, the high-quality UV–optical lag-spectrum constructed by \citet{fausnaugh2016} was well reproduced using the same modeling approach by \citet{kammoun2021}.

\begin{figure}
\centering
\includegraphics[width=0.45\textwidth]{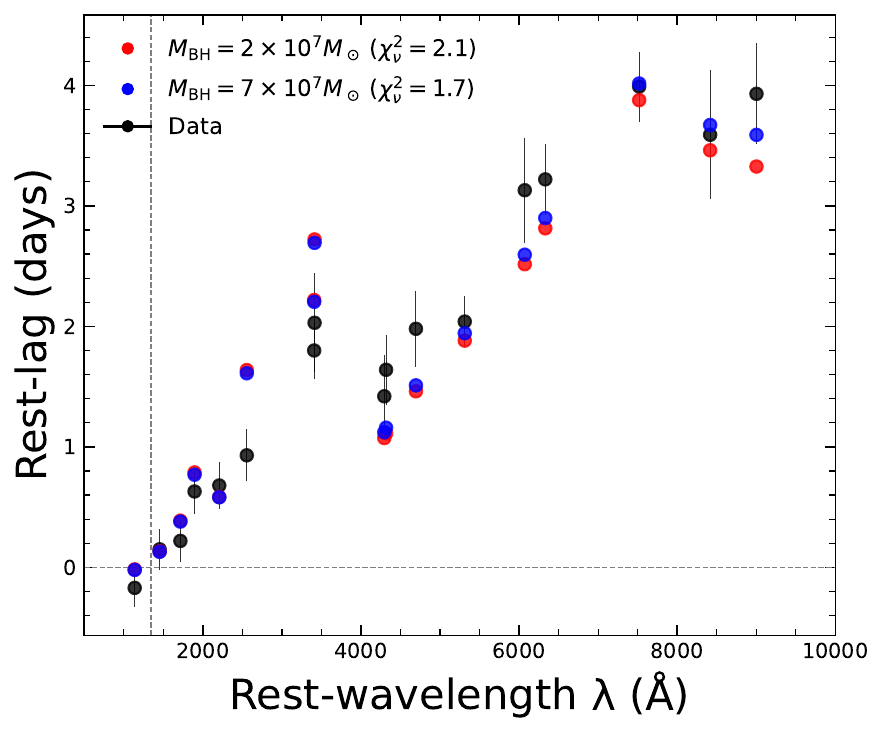}
\includegraphics[width=0.45\textwidth]{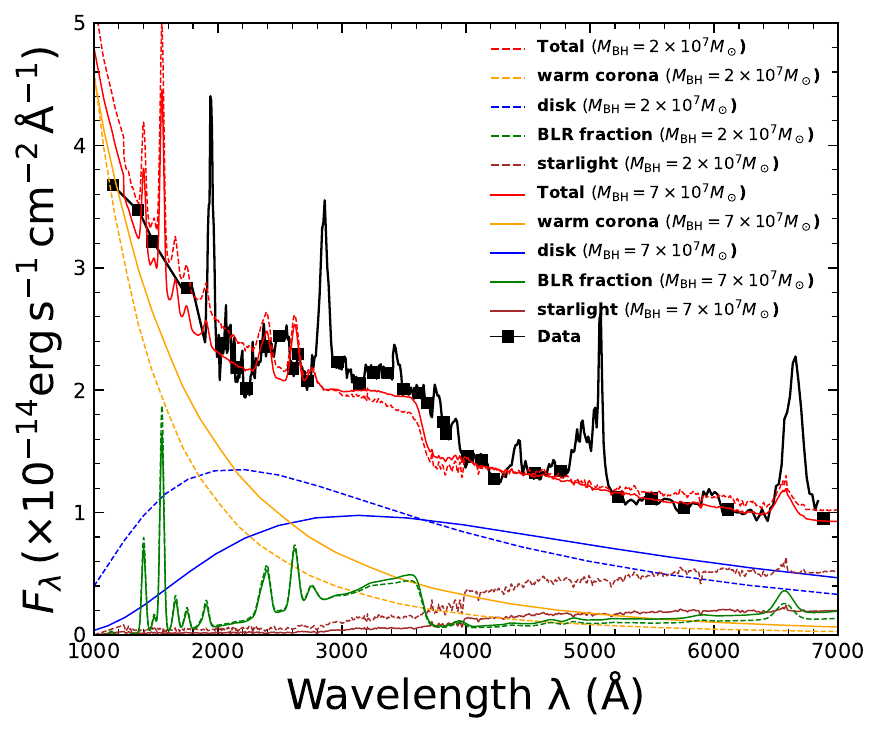}
\caption{UV-optical lag-spectrum and SED of NGC~5548. Top: the black circular points with error bars show the observed rest-frame inter-band time delays. The inter-band time delays recovered from the accretion disk combined with BLR reprocessing model for two $M_{\mathrm{BH}}$ values are shown by red and blue circular points. Vertical and horizontal dotted lines indicate the rest-frame reference wavelength and zero rest-frame lag, respectively. Bottom: the UV-optical spectrum, where the black solid line represents the observed data, and the black squares mark the selected continuum points. The individual model components are shown in different colors, while the total best-fit models are indicated by red lines for $M_{\mathrm{BH}} = 2 \times 10^7 \, M_{\odot}$ and $M_{\mathrm{BH}} = 7 \times 10^7 \, M_{\odot}$, shown as dashed and solid lines, respectively.
}
\label{fig:frado_lgspc}
\end{figure}

\subsection{Assumptions in X-ray$-$RM modeling}

 The model we employ incorporates several advanced features, including GR effects and a self-consistent treatment of hard X-ray reprocessing by the accretion disk. However, it does not capture all relevant physical processes. In particular, the irradiation is modeled using a lamp-post geometry, while the hard X-ray emission is assumed to originate from a spherical hot corona.

A key parameter in this context is the compactness parameter, a dimensionless combination of the corona radius (when approximated as a sphere) and its luminosity \citep{guilbert1983,svensson1984}. The interaction of soft photons with hot electrons additionally depends on the electron temperature (for a thermal electron distribution) and the optical depth, as pair creation can act as an important cooling mechanism, preventing the corona from becoming too compact \citep{bisnovatyi_kogan1971,svenson_zdziarski1994}. Spectral analysis allows the evaluation of some coronal plasma parameters \citep[e.g.,][]{tortosa2018}, and relatively recent observational studies suggest that corona sizes are of the order of $3$–$10 \, r_g$ \citep{fabian2015}, highlighting the significant role of pair creation.

 Thus, the point-like lamp-post approximation represents a simplifying assumption. In the model, the radius of the hot corona is explicitly calculated and is typically of the order of half of the corona height; however, the disk irradiation is still treated as originating from a point-like source. Spectral data alone place only weak constraints on the detailed geometry of the corona, since the reflected emission from the underlying accretion disk is relatively insensitive to the exact spatial extent and shape of the irradiating region. Robust constraints on the corona geometry therefore require extremely high-quality data \citep[e.g.,][]{szanecki2020,feng2025,nekrasov2025}. Consequently, the adoption of the lamp-post approximation is not expected to significantly affect the results presented here, if the hot corona indeed has a spherical geometry.

  However, this assumption may not hold in all cases. If the corona is, in fact, part of a (not very strong) jet \citep{henri1991} with standing or propagating shocks, we can envision two distinct regions of relatively high dissipation. In such a scenario, a vertically extended corona could help resolve the apparent discrepancy in inferred coronal heights, since X-ray and UV–optical observations probe different reprocessing regions. Specifically, X-ray reflection arises within a few $r_g$, where the lower portion of the corona is likely to dominate, whereas UV–optical emission is produced at distances of tens to hundreds of $r_g$, where the upper regions of the corona may play a more significant role.

  An additional challenge lies in incorporating variations in the hard X-ray spectral slope into the model in a fully self-consistent manner. Although this could, in principle, be achieved by adopting a time-dependent spectral slope directly from the observations, such an approach requires reliable estimates of the hard X-ray spectral slope obtained through a dedicated reanalysis of the X-ray data. Since no such analysis was performed in the present work, this aspect has not been implemented here; however, we plan to address it in future studies.

The role of the warm absorber is also not negligible. \citet{Turner2009} highlighted a significant contribution of absorption (warm absorber) on top of reflection, potentially affecting its properties. While \citet{Panagiotou2019} concluded that in unobscured sources the reflection primarily originates from the disk. However, NGC 5548 does show traces of variable obscuration. Moreover, the mechanism responsible for hard X-ray positive lags may be coupled to soft negative flux variations, further complicating the modeling. If this absorption originates from a warm absorber along the line of sight with density $\sim 10^9$ cm$^{-3}$, the corresponding recombination timescale of order one hour could further influence the observed variability characteristics. Furthermore, the variable absorber may also not cover the entire source, as for example envisioned in Figure~14 of \citet{wildy2021}, which would introduce additional modifications to the measured time delays.

%Taken together, these issues suggest that the available X-ray data may not be fully adequate for the level of analysis performed. More generally, reliable time-lag estimation requires a substantial amount of high-quality data to ensure unbiased lag measurements with uncertainties that accurately reflect the true variance. In particular, lags should be computed only at frequencies where the observed coherence is sufficiently high. To strengthen the reliability of the results and better test the proposed models, a re-estimation of the X-ray time lags using all available archival data may therefore be necessary.

\subsection{Assumptions in UV-optical$-$RM modeling}
\label{ss:frado}

 A key assumption in the present study is that the contribution from the BLR is neglected. However, there is growing evidence that the BLR contributes to the UV-optical bands not only through strong emission lines but also via a reprocessed continuum \citep[e.g.][]{Korista_Goad_2001,netzer2022,jaiswal2023}.

If this additional BLR component is taken into account, the total observed delay would include both the intrinsic disk response and an extra contribution from the BLR. As a result, the disk’s intrinsic lag could be shorter than currently inferred, which in turn may lead to a lower estimate of the corona height from model fits. At present, the model employed in this work does not incorporate such a BLR contribution, and therefore this effect cannot yet be explored within our framework.

%The discrepancy in the inferred corona height is easier to rationalize, as the commonly adopted point-like approximation for the hard X-ray source is physically unrealistic. The hot corona is required to produce the observed hard X-ray power law, which is most commonly explained by Comptonization. This process requires an extended region filled with hot plasma, where soft photons from the accretion disk are intercepted and upscattered to higher energies.

%\subsection{Consistency check with an independent model for the discrepancy in $M_{\text{BH}}$ and $\dot{m}$}

 Hence, to investigate the discrepancy between the $M_{\text{BH}}$ and $\dot{m}$ inferred from the X-ray$-$RM and UV–optical$-$RM modeling, together with their respective SED fits in NGC~5548, we apply an additional independent methodology. Specifically, we adopt the simultaneous UV–optical lag-spectrum and broadband SED fitting approach described by \citet{jaiswal2025}. This framework incorporates a standard accretion disk, an inner hot flow, a warm corona covering part of the inner disk, lamp-post irradiation of the outer disk, and a radiation-pressure regulated BLR based on the Failed Radiatively Accelerated Dusty Outflow (FRADO) model \citep{czhr2011, naddaf2021, naddaf2022}, which accounts for BLR contamination \citep[e.g.,][]{Korista_Goad_2001, lawther2018, netzer2022}. However, this model does not include GR effects or the albedo contribution from the disk, and it instead assumes complete thermalization of the incident X-ray flux. As a result of these simplifications, the model is not well suited for describing X-ray reverberation phenomena.

The goal is to assess which $M_{\text{BH}}$ estimate, either that derived from X-ray SED and X-ray$-$RM analysis or that obtained from UV–optical lag-spectrum and SED fitting is more consistent when confronted with an independent modeling strategy, since such a large discrepancy in both  $M_{\text{BH}}$ and $\dot{m}$ is physically unexpected.

The top panel of Figure \ref{fig:frado_lgspc} shows the model-recovered lag-spectra for two trial black hole masses, $M_{\text{BH}} = 2 \times 10^7 \, M_{\odot}$ and $M_{\text{BH}} = 7 \times 10^7 \, M_{\odot}$, overlaid on the observed time-delay measurements. As part of the same procedure, we simultaneously fit the UV-optical SED of NGC~5548 from \citet{mehdipour2015}. The bottom panel of Figure~\ref{fig:frado_lgspc} presents these SED fits, where the best-fit models for the two masses are shown as red dashed and red solid curves, respectively. The black squares mark the selected continuum points extracted from the observed spectrum (black line), and all contributing model components are displayed to illustrate how the total SED is constructed. In this fitting, the corona height and accretion rate were allowed to vary freely. The corresponding best-fit parameters are reported in Table \ref{tab:mod_uvop}.

We find that the model provides a better joint fit to both the UV–optical lag-spectrum and the SED for $M_{\text{BH}} = 7 \times 10^7 \, M_{\odot}$, resulting in $\chi^2_{\nu} = 1.7$ ($p$-value = 0.038) for the lag-spectrum and $\chi^2_{\nu} = 0.84$ for the SED, compared to $\chi^2_{\nu} = 2.1$ ($p$-value = 0.007) and $1.97$, respectively, for $M_{\text{BH}} = 2 \times 10^7 \, M_{\odot}$. These results further indicate that NGC~5548 favors a larger black hole mass of $M_{\text{BH}} = 7 \times 10^7 \, M_{\odot}$ together with a smaller accretion rate of $\dot{m} \sim 0.014$ and a lamp-post height of $h = 18.6 \, r_g$. It is worth noting that \citet{jaiswal2025} fitted the same lag-spectrum and SED assuming $M_{\text{BH}} = 5 \times 10^7 \, M_{\odot}$. They obtained broadly similar results, with $\dot{m} \sim 0.017$ and a larger corona height of $h = 48.3 \, r_g$. Importantly, these estimates are broadly consistent with the constraints derived from the {\tt KYNXiltr} modeling of the UV–optical lag-spectrum and the broadband SED, once the dispersion associated with different black hole spin configurations is taken into account. Consequently, explaining the low $M_{\mathrm{BH}}$ and high $\dot{m}$ solution inferred from the X-ray SED and Fourier-resolved X-ray$-$RM analysis appears highly unlikely. 

 Notably, the model of \citet{Kammoun2024} incorporated GR effects but excluded contributions from the BLR, whereas the model of \citet{jaiswal2025} included the BLR but did not account for GR effects. Despite these differences, both studies favored larger black hole masses and corona heights, together with relatively lower accretion rates, which successfully reproduce the corresponding SED shapes and UV–optical time delays across multiple wavelength bands. Therefore, further work on X-ray$-$RM is clearly warranted.

% recombination timescale: 1/(n⋅2.6×10−13)

\subsection{Future prospects}

 The potential limitations affecting the X-ray results used in the present study prevent us from drawing firm conclusions that the observed X-ray$-$UV/optical discrepancy necessarily reflects differences in the underlying assumptions of the reverberation models. At the same time, the assumptions adopted in our modeling of the Fourier-resolved lags may themselves contribute to the apparent X-ray$-$UV/optical inconsistency.

The essence of the X-ray$-$RM model is the assumption that the X-ray source emits a flux with a fixed spectral shape, a power law with photon index $\Gamma$ that does not vary with time. This isotropic emission is partially observed directly by the observer and partially illuminates the accretion disk. The subsequent reprocessing involves selective absorption and re-emission, which depend on the ionization state of the disk, calculated self-consistently from the disk surface density and the incident radiation. Consequently, a spectral evolution arises entirely from the reprocessing of radiation by progressively more distant disk regions. The disk’s internal dissipation is assumed to be constant, hence its variability is solely a response to irradiation. Moreover, the modeling does not account for hard X-ray lags. Additionally, the model assumes the absence of a warm absorber along the line of sight that could respond to changes in the lamp luminosity.

Each of these assumptions can potentially influence the results. The corona is compact, and any internal changes can occur rapidly and may be coupled to variations in the soft photon flux from the innermost regions of the flow. Modeling such changes is beyond the scope of this study, given the limited understanding of the variability mechanisms in this complex region. 

\section{Summary}

\label{sec:sum}

  We modeled the X-ray SED, the Fourier-resolved  lag–energy and lag–frequency spectra from X-ray$-$RM, and the inter-band lag-spectrum from UV-optical continuum$-$RM of NGC~5548 within a unified and physically motivated reflection-based framework. Despite adopting a consistent modeling strategy across all wavelength regimes, we were unable to identify a single set of physical parameters that can simultaneously reproduce both the spectral and timing properties of the source.

The Fourier-resolved X-ray$-$RM data preferentially indicate a relatively small black hole mass, $M_{\text{BH}} \sim 2-3 \times 10^7 \, M_{\odot}$, together with a higher accretion rate and a compact corona. This parameter combination also yields an improved description of the X-ray SED. In contrast, the UV–optical continuum$-$RM data and broadband SED consistently favor a larger black hole mass, $M_{\text{BH}} \sim 5-7 \times 10^7 \, M_{\odot}$, lower accretion rates, and a more spatially extended irradiating region. While discrepancies in the inferred corona height can be partly attributed to the simplified lamp-post geometry and to the fact that X-ray and UV–optical reverberation probe markedly different spatial scales, the inconsistency in the inferred $M_{\text{BH}}$ and $\dot{m}$ represents a more fundamental challenge to achieving a fully self-consistent model.

In addition, we identify several potential limitations in the X-ray$-$RM data used in \citet{kara2016}. The lag-energy and lag-frequency spectra are derived from a single XMM-Newton observation characterized by modest intrinsic variability, and uncertain coherence at high frequencies. Together, these factors may bias the inferred Fourier-resolved lags and, consequently, the physical parameters derived from X-ray$-$RM modeling. Moreover, the lack of a consistent treatment of intrinsic X-ray absorption and warm absorbers in the construction of the X-ray SED data may contribute to the significant deviations observed in the model fits.  A re-analysis based on the full archival X-ray dataset, incorporating coherence-based frequency selection and a consistent treatment of absorption, is therefore essential to obtain more robust and reliable constraints.

 We conclude that the most critical next step is to fully exploit the available X-ray data for this source by analyzing a larger dataset and making use of its full potential. At the same time, future progress will also require extending the current modeling framework by incorporating additional physical ingredients that are presently missing.

%\citet{arevalo2008} - they do not have time delays despite using Fourier-resolved results...

\begin{acknowledgements}
 We thank the referee for comments and suggestions. We also thank Piotr T. \. Zycki for helpful discussions. 
This project has received funding from the European Research Council (ERC) under the European Union’s Horizon 2020 research and innovation program (grant agreement No. [951549]). The Czech-Polish Mobility program of the two Academies of Sciences, titled
“Appearance and dynamics of accretion onto black holes”, is greatly appreciated.
\end{acknowledgements}

\bibliographystyle{aa}
\bibliography{main}

\begin{appendix}
\label{ss:apendix}
\onecolumn

\section{Fourier resolved- lag dependence on black hole mass in the model}
\label{ss:ap_A}

The lag–energy spectrum was computed in a fixed Fourier frequency band of $(0.4-3) \times 10^{-4}$ Hz \citep{kara2016} , and this same band was adopted throughout our analysis. Because larger black holes have longer characteristic timescales, their variability power shifts to lower Fourier frequencies, whereas lower-mass AGNs exhibit variability at higher frequencies. Consequently, a fixed Fourier frequency band samples different regions of the transfer function for different $M_{\text{BH}}$.

For low $M_{\text{BH}}$, the characteristic frequencies are high, thus the chosen frequency band may fall below the break frequency, where propagation lags dominate or partially overlap the reverberation regime. In this case, the band probes relatively slow variability for that mass, resulting in larger measured lags. For high $M_{\text{BH}}$, however, the same frequency band may lie in the high-frequency tail of the transfer function, where the response is weak or even beyond the coherence limit; the band then probes variability faster than physically relevant for that mass, producing smaller lags (see Figure~\ref{fig:model_fit}). Thus, the Fourier-resolved lag amplitude decreases systematically with increasing $M_{\text{BH}}$ when the frequency band is held fixed.

\begin{figure}[h!]
\centering
\includegraphics[width=0.45\textwidth]{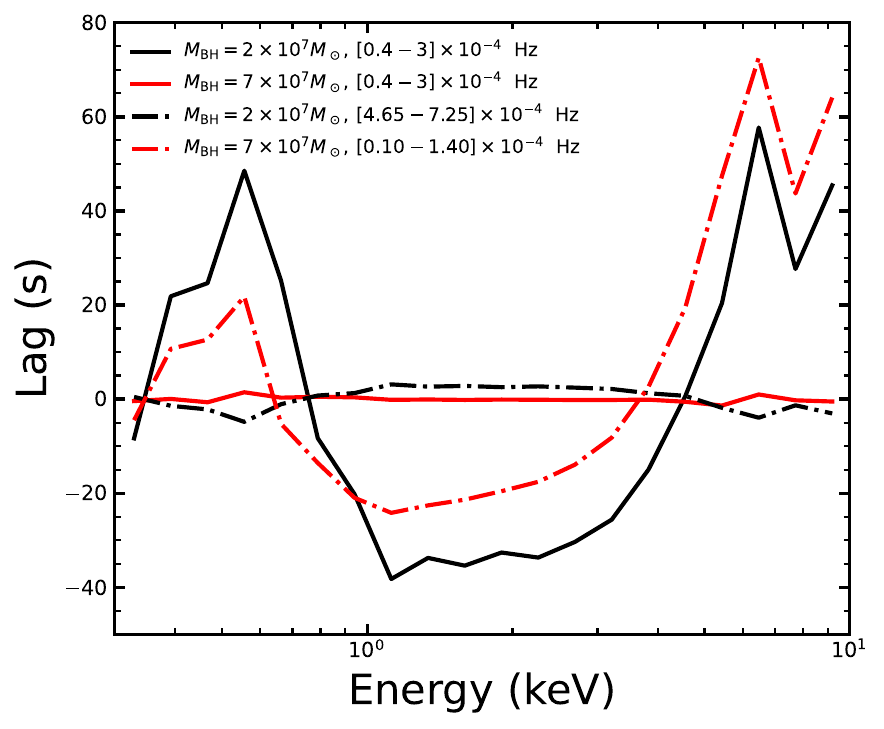}
\caption{Model-fitted lag-energy spectra for different $M_{\mathrm{BH}}$, computed with $\dot{m}=0.05$, a fixed height of $h = 5 \, r_g$, and all other parameters held constant except the Fourier frequency band. The black solid line shows the model for $M_{\mathrm{BH}} = 2 \times 10^7 \, M_{\odot}$ using the $(0.4-3) \times 10^{-4}$ Hz band. The red solid line shows the same frequency band applied to $M_{\mathrm{BH}} = 7 \times 10^7 \, M_{\odot}$. The black dot-dashed and red dot-dashed lines show the models for $M_{\mathrm{BH}} = 2 \times 10^{7} \, M_{\odot}$ and $M_{\mathrm{BH}} = 7 \times 10^{7} \, M_{\odot}$, respectively, computed using the mass-scaled frequency bands of $(4.65 - 7.25)\times 10^{-4} \, \mathrm{Hz}$ and $(0.10 - 1.40)\times 10^{-4} \, \mathrm{Hz}$. 
}
\label{fig:ener_mas_exp}
\end{figure}

To illustrate this, we computed lag–energy spectra for a lower black hole mass, $M_{\text{BH}} = 2 \times 10^{7} \, M_{\odot}$, and for a higher mass, $M_{\text{BH}} = 7 \times 10^{7} \, M_{\odot}$, using both the original Fourier band of $(0.4 - 3)\times 10^{-4}$ Hz and frequency bands scaled upward and downward by the mass ratio of 3.5. The corresponding mass-scaled frequency ranges are $(4.65 - 7.25)\times 10^{-4}$ Hz and $(0.10 - 1.40)\times 10^{-4}$ Hz, respectively. As shown earlier, comparing the two masses within the same fixed original Fourier band (solid lines) results in larger lag amplitudes for the lower-mass black hole. However, when the Fourier band is shifted upward for the lower-mass model, or downward for the higher-mass model, the resulting lag–energy spectra (black dot–dashed and red dot–dashed lines, respectively) align closely with those of the higher- or lower-mass models computed using the original frequency band (red and black solid lines) as shown in Figure~\ref{fig:ener_mas_exp}.

In summary, the apparent decrease in Fourier-resolved lag amplitude with increasing  $M_{\text{BH}}$ arises naturally when a fixed frequency band is used, because the same band probes progressively faster and less responsive regions of the transfer function for higher-mass black holes. Adjusting the Fourier frequency band in proportion to  $M_{\text{BH}}$ restores consistency between the modeled lag–energy spectra, demonstrating that the mass dependence of the observed lags in the lag-energy spectrum is primarily a consequence of the characteristic timescale scaling with black hole mass rather than intrinsic changes in the reverberation response. However, since our simulations were performed using the same approach as in the data analysis of \citet{kara2016}, the discrepancy between the model predictions and the observed data persists.

\end{appendix}

\end{document}